\documentclass[%
twocolumn,superscriptaddress,
 amsmath,amssymb,
 aps,
prl,
]{revtex4-2}
\usepackage{mathrsfs,amsmath,amssymb,amsthm}
\usepackage{amssymb,fmtcount}
\usepackage{graphicx,epsfig,latexsym,overpic,amssymb,color}
\usepackage{threeparttable}
\preprint{\today}

\begin{document}
\title{
 Speed of Sound and Phase Transitions in Neutron Stars Indicated by the Thick Neutron Skin of $^{208}$Pb
}
\author{Manjia Chen}
\affiliation{
State Key Laboratory of Nuclear Physics and Technology, School of Physics,
Peking University, Beijing 100871, China
}
\author{Dawei Guan}
\affiliation{
State Key Laboratory of Nuclear Physics and Technology, School of Physics,
Peking University, Beijing 100871, China
}
\author{Chongji Jiang}
\affiliation{
State Key Laboratory of Nuclear Physics and Technology, School of Physics,
Peking University, Beijing 100871, China
}
\author{Junchen Pei}\email{peij@pku.edu.cn}
\affiliation{
State Key Laboratory of Nuclear Physics and Technology, School of Physics,
Peking University, Beijing 100871, China
}
\affiliation{
Southern Center for Nuclear-Science Theory (SCNT), Institute of Modern Physics, Chinese Academy of Sciences, Huizhou 516000,  China
}

\begin{abstract}
The speed of sound is a novel probe of equation of state and phase transitions
in dense cores of neutron stars.
Recently nuclear experiments extracted a surprising thick neutron skin of
$^{208}$Pb, causing tensions to reproduce the tidal deformability in gravitational-wave observations.
This work finds that exotic structures in the speed of sound with a small softening slope followed by a steep-rising peak are required to
reconcile the thick neutron skin of $^{208}$Pb with  astronomical observations of neutron stars. 
Furthermore, the peak of speed of sound is narrowly constrained around two times the nuclear saturation density  with the thick neutron skin.
Consequently early and strong first-order phase transitions are comparatively more favorable.

\end{abstract}
\maketitle

\emph{Introduction.}---
Dense nuclear matter is of fundamental interests to study strong interactions and the
phase transition from nuclear matter to quark matter.
The recent advent of astronomic observation facilities such as LIGO/Virgo~\cite{gw170817a,gw170817} and NICER~\cite{j0030-a,j0030-b} has  brought
unprecedented opportunities to explore the dense cores of neutron stars.
However, it is unlikely to constrain the equation of state (EoS) of dense nuclear matter
 by astronomical observations alone.
In nuclear physics, EoS around the nuclear saturation density ($n_{sat}$$\approx$0.16 fm$^{-3}$) is largely known,
providing a stepping stone for extrapolations to higher densities.
It is possible to constrain  EoS up to 5 times $n_{sat}$ with  heavy-ion collision experiments but restricted to nearly-symmetric nuclear matter~\cite{collision}.
The first-principle  descriptions of nuclear matter from chiral effective field theory ($\chi$EFT) also
suffer from large uncertainties towards higher densities~\cite{sound-xft}, due to
 ambiguous many-body interactions.
In addition, strongly interacting matter at extremely high densities $\thicksim$40$n_{sat}$
can be reliably calculated by the perturbative quantum chromodynamics (pQCD)~\cite{pqcd}.
Therefore the study of nuclear matter at a few times  $n_{sat}$, i.e., the possible scenario of phase transition,  by applying constraints from multiple messengers
is a destined choice.

It is always intriguing to connect properties of finite nuclei and observations of neutron stars.
Among existing nuclear observables to constrain EoS,
it is worthy to mention the recent PREX and PREX-II experiments on measuring the neutron skin of $^{208}$Pb through parity violation in electron scattering~\cite{prex,prex2}.
The extracted neutron skin thickness is $R_{skin}^{208}$=0.283$\pm$0.071 fm~\cite{prex2}, which is abnormally large
and presents a serious challenge to current nuclear theories.
Advanced ab initio calculations of $^{208}$Pb predict a neutron skin thickness about 0.14$-$0.20 fm~\cite{bshu}.
The comprehensively inferred $R_{skin}^{208}$ is 0.17 fm by considering constraints of astrophysical data~\cite{skin-2}.
The large $R_{skin}^{208}$ leads to a large radius and large tidal deformability of neutron stars of 1.4 solar mass ($M_\odot$),
causing tensions with  gravitational-wave observations in the GW170817 event~\cite{skin-1}.
The experimental neutron skin of $^{48}$Ca~\cite{crex} is slightly less than theoretical estimations and
 is not an imperative issue.

The behaviors of speed of sound in neutron stars have attracted great interests recently~\cite{sound1,sound-2}.
The speed of sound $c_s$ is defined as
\begin{equation}
c_s^2=  \frac{\partial p(\epsilon)}{\partial \epsilon }
\nonumber
\end{equation}
where $p$ is the pressure and $\epsilon$ is the energy density.
Therefore,  EoS can be determined via an integral of $c_s^2$
without details of nuclear interactions~\cite{sound1}.
The speed of sound $c_s$ in the pQCD regime is reliably calculated to be $1/\sqrt{3}$ of the speed of light~\cite{pqcd},
providing a novel constraint on EoS at extremely high densities. 
Around the saturation density, the speed of sound is defined by the incompressibility of nuclear matter~\cite{sly4}.
The speed of sound between these two regimes is essentially unknown.
The allowed $c_s$ from exhaustive searches and the possibilities of hybrid neutron stars have been extensively explored~\cite{sound1,sound-2}.
It is largely believed that with increasing
densities, $c_s$ should first rise to surpass $1/\sqrt{3}$ and then
decrease and finally approach  $1/\sqrt{3}$~\cite{sound-2}.
The speed of sound is also a direct signature of phase transitions, in which
various nonsmooth structures such as bumps, spikes, step functions, plateaus, and kinks are possible~\cite{sound-3}.

The aim of this work is to reveal the impacts of the abnormal thick neutron skin of $^{208}$Pb
on the speed of sound and EoS of neutron stars.
There were extensive studies of consequences of the thick neutron skin of $^{208}$Pb~\cite{bshu,skin-1,skin-2,prx-skin,skin-4,skin-5,skin-6,skin-7,lwchen},
or structures of speed of sound in neutron stars~\cite{sound-xft,sound1,sound-2,sound-3,sound-4,sound-5,sound-6,sound-7,sound-8,sound-9} separately.
This work performs the first combined study and finds that exotic structures
in speed of sound are required to reconcile the thick neutron skin with astronomical observations of neutron stars.

\emph{Methods.}---
Firstly a set of new Skyrme density functionals are obtained by optimizing the properties
of finite nuclei and infinite nuclear matter around $n_{sat}$.
In particular, the neutron skin thickness of $^{208}$Pb and $^{48}$Ca are included.
For finite nuclei, the binding energies and charge radii of 50 nuclei in the nuclear landscape are adopted,
as described in~\cite{xiong}.
To better describe finite nuclei and neutron stars simultaneously,
a higher-order density-dependent term is added~\cite{xiong}.
The fitting procedure is realized using the simulated annealing method.
There is a competition in the fitting between properties of finite nuclei and the thick neutron skin.
Hence the weights of the neutron skin thickness are varied to
obtain different parameter sets.
The EoS of nuclear matter is constructed based on obtained density functionals as well as
the $npe\mu$  $\beta$-equilibrium.

Next EoS at densities higher than 1.2$n_{sat}$ is constructed from the speed of sound without
any details of nuclear interactions, according to Ref.~\cite{sound1}.
The number density $n$ can be obtained by the integral of sound speed in terms of
the chemical potential $\mu$ from a starting density $n_1$, as $n(\mu)=n_1{\rm exp}(\int_{\mu_1}^{\mu} \frac{d\mu'}{\mu' c_s^2})$.
Then the pressure $p(\mu)$ is obtained by an additional integration of the number density $n(\mu)$
 as $p(\mu)=p_1+\int_{\mu_1}^{\mu} d\mu' n(\mu)$.
Note that $\mu_1$, $p_1$ correspond to values at the starting density $n_1$ to match EoSs given by density functionals.
Since the sound speed is essentially unknown at intermediate densities,
we adopt analytical curves to connect the starting $n_1$ and the peak location of  $c_s^2$.
In the first part from $n_1$ to the peak, we choose a monotonically increasing 4th-order polynomial curve after extensive tests.  
The polynomial coefficients are determined by two endpoints, in which both $n_1$ and the peak are varying.
In the second part, the speed of sound is likely to decrease and finally matches $1/\sqrt{3}$.
The second part only slightly affects massive neutron star observations.
The possible zero $c_s^2$ related to the first-order phase transition is also allowed in the second part.
The peak locations of  $c_s^2$ are focused and heavily searched in this work, and the total computational costs are much less than the piecewise linear interpolation search in Refs.~\cite{sound1,sound-2}.

The astronomical observations of neutron stars are calculated by solving the well-known
Tolman-Oppenheimer-Volkoff (TOV) equations.
Based on EoS from density functionals and the speed of sound together, TOV equations are solved with varying initial densities at the neutron star center
to obtain the mass-radius relationship.
The tidal deformability $\Lambda$ as a function of the neutron star mass can also be calculated by solving the TOV equation,
as described in our previous work~\cite{jiang}.

\begin{figure}[t]
\centering
\includegraphics[width=0.46\textwidth]{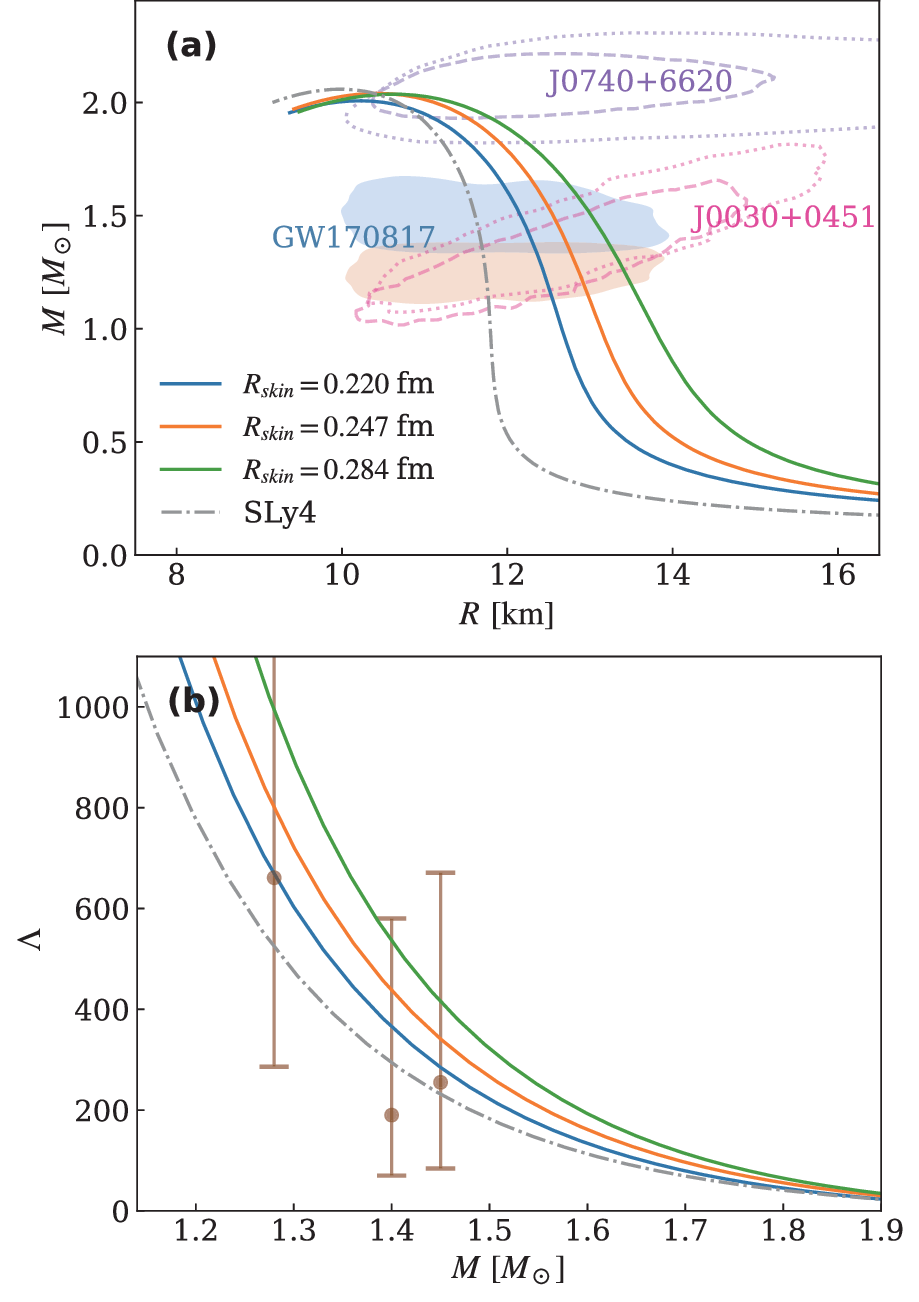}
\caption{
Calculated neutron star observations with different density functionals associated with varying
neutron skin thickness of $^{208}$Pb. (a) the mass-radius curves; (b) 
the calculated tidal deformability $\Lambda$ and the GW170817 inferences~\cite{gw170817}.
                           \label{FIG1}
}
\end{figure}

\emph{Results.}---
To implement constraints from the neutron skin of $^{208}$Pb, we obtain three newly optimized
 Skyrme density functionals with different fitting weights, which correspond to
$R_{skin}^{208}$=0.22, 0.247, 0.284 fm, respectively.
The resulting mass-radius relations of neutron stars are shown in Fig.\ref{FIG1}.
For comparison, results of SLy4~\cite{sly4} are also shown, which corresponds to $R_{skin}^{208}$=0.16 fm.
We can see that, with an increasing $R_{skin}^{208}$, the radius and tidal deformability of a 1.4$M_\odot$ neutron star also increase, while
 the maximum masses all are around 2.0$M_\odot$.
For results corresponding to $R_{skin}^{208}$=0.284 fm, which is close to the PREX-II experiment,
the associated radius and tidal deformability of a 1.4$M_\odot$ neutron star are in tension with the GW170817 observations, in which
$\Lambda=190_{-120}^{+390}$~\cite{gw170817}. Such tension is more serious in the relativistic mean-field framework~\cite{skin-1}.
All density functionals can not reproduce the massive neutron star in PSR J0740+6620 from NICER and XMM-Newton observations~\cite{j0740-a,j0740-b}.

\begin{figure*}[t]
\centering
\includegraphics[width=0.95\textwidth]{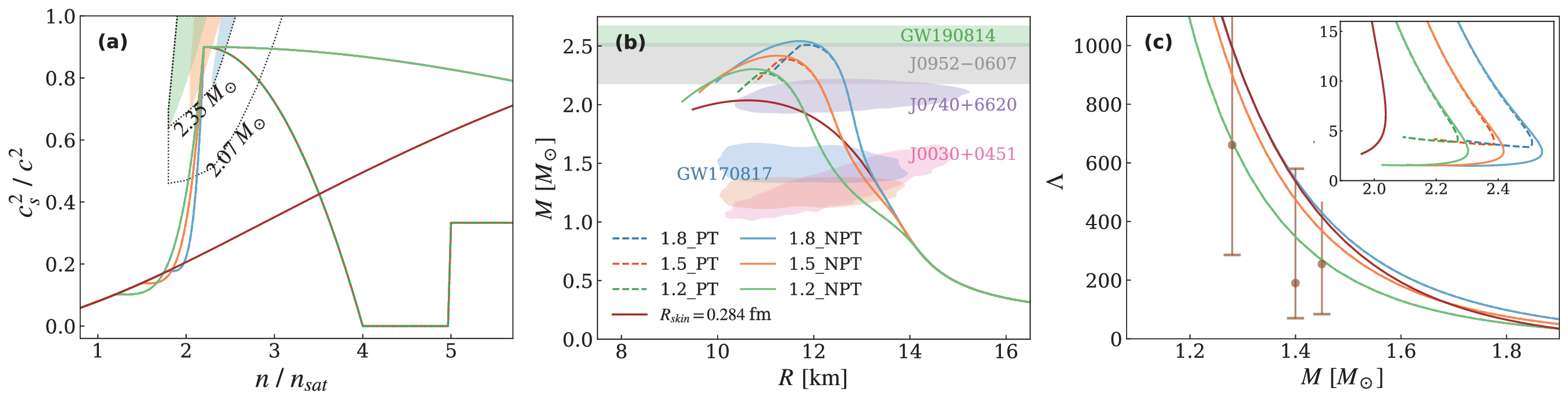}
\caption{
Neutron star observations based on EoS from the combination of the density functional of the thick neutron skin and different speed of sound.
(a) exotic structures of speed of sound with constrained peak regions, in which constraints of maximum masses above 2.35 and 2.07$M_\odot$ are shown; (b) the mass-radius curves corresponding to different $c_s^2$ in (a),
in which structures with (PT) and without (NPT) the first-order phase transition are compared; (c) the tidal deformability $\Lambda$ associated with
different speed of sound in (a).
 \label{FIG2}
}
\end{figure*}

\begin{figure}[b]
\centering
\includegraphics[width=0.49\textwidth]{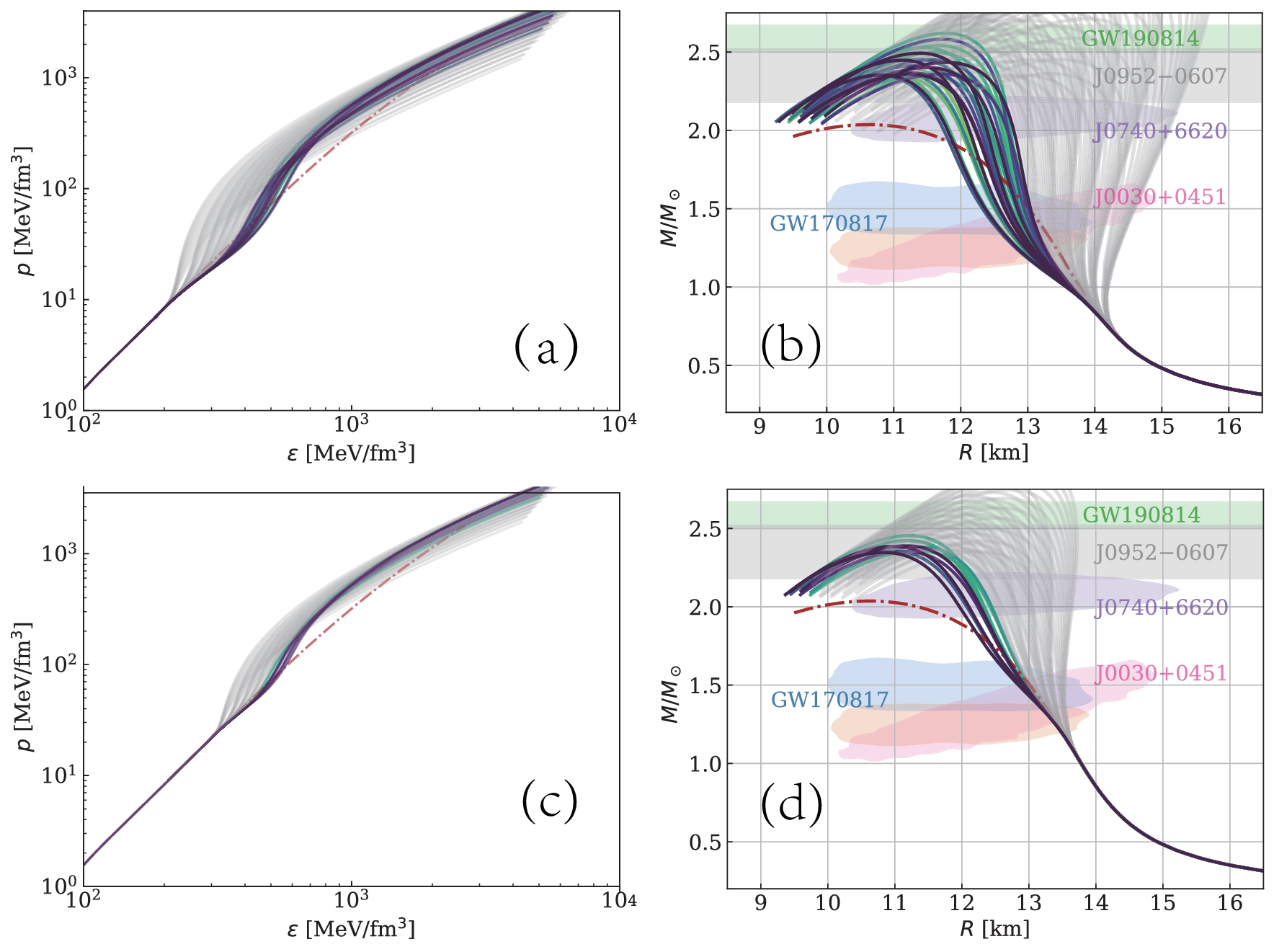}
\caption{
EoS and neutron star observations by searching the speed of sound associated with the thick $R_{skin}^{208}$,
in which the results beyond $\Lambda$=519 at 1.4$M_\odot$ are shown in gray color. 
(a) the search of EoS with the starting density $n_1$=1.2$n_{sat}$;
(b) the resulting mass-radius curves with $n_1$=1.2$n_{sat}$;
(c)  EoS with  $n_1$=1.8$n_{sat}$; (d) mass-radius curves with $n_1$=1.8$n_{sat}$.
\label{FIG3x}
}
\end{figure}

To reconcile the thick neutron skin with astronomical observations of neutron stars,
the varying speed of sound is constructed starting from $n_1$=1.2, 1.5, 1.8$n_{sat}$.
After heavy searches, we find that an early softening speed of sound is required
to reduce the radius and tidal deformability of a 1.4$M_\odot$ neutron star.
Then a steep-rising speed of sound is required to obtain the reported maximum mass.
With earlier modifications of the speed of sound, there are larger rooms to adjust $c_s$
to reproduce astronomical observations.
Similarly, the starting density $n_1$ is adopted as 1.1$n_{sat}$ in Refs.~\cite{sound1,sound-2}.
Calculations show that neutron star observations are sensitive to the speed of sound below 3$n_{sat}$
and  are barely dependent on details of $c_s^2$ at higher densities.
It can seen that, by employing exotic structures in speed of sound, the observations of neutron stars in
GW170817~\cite{gw170817}, J0030+0451~\cite{j0030-a,j0030-b}, J0740+6620~\cite{j0740-a,j0740-b} can now be described simultaneously.
The slightly softening sound speed at the beginning is crucial
to satisfy the tidal deformability of neutron stars in GW170817.
In Ref.~\cite{skin-2},  the statistical analysis also finds that a softening of speed of sound is possible
when the slope of symmetry energy is larger than 100 MeV. 
Moreover, the reported heaviest neutron star of 2.35$M_\odot$ in pulsar J0952-0607~\cite{j0952}  would inevitably invoke a steep-rising speed of sound.
In Fig.\ref{FIG2}(b), a possible neutron star of 2.6$M_\odot$ in the GW190814 event~\cite{gw190814} can be reached with $n_1$=1.8$n_{sat}$, but it violates the tidal deformability constraints.  
The peak in Fig.\ref{FIG2}(a) is  at 2.2$n_{sat}$, and by shifting the peak to 1.9$n_{sat}$, a 2.6$M_\odot$ neutron
star can also be reached with $n_1$=1.2$n_{sat}$. However,  it is difficult to reach 2.6$M_\odot$ by applying
the tidal deformability constraint at 1.4$M_\odot$ with the thick $R_{skin}^{208}$, as shown in Fig.\ref{FIG3x}.
Note that whether it is a massive neutron star or a light black hole is still in disputation~\cite{disp-1}.

One crucial question is to search for the allowed region of peak locations in speed of sound. 
For the thick $R_{skin}^{208}$, the exotic structures in sound speed are inevitable  even for a 2.0$M_\odot$ neutron star based on heavy searches.
In Fig.\ref{FIG2}(a), the allowed region of the peak locations with constraints on maximum masses at 2.07$M_\odot$~\cite{j0740-a,j0740-b} and 2.35$M_\odot$ ~\cite{j0952} are shown.
The left boundaries are determined by the tidal deformability $\Lambda$=519 at 1.4$M_\odot$, as suggested by a comprehensive study~\cite{519}, which 
is a slightly more stringent constraint compared to the observational upper limit at 580~\cite{gw170817}.
It can be seen that with combined constraints of 2.35$M_\odot$,  $\Lambda$=519, and the thick $R_{skin}^{208}$, the allowed region of the peak locations is rather shrunk. 
With different $n_1$, the corresponding allowed peak area is rather small, as indicated by different shadow colors in Fig.\ref{FIG2}(a). 
The searches show that $n_1$=1.8$n_{sat}$ is too late to describe GW170817 observations.

The possible first-order phase transitions with zero speed of sound at 4$n_{sat}$ are also displayed in Fig.~\ref{FIG2}.
The observations of such sharp phase transitions are different from those of smooth transitions only in neutron stars around the maximum masses.
We see that there are possible two neutron stars with similar masses but different radii, called twins,
as an evidence of phase transition.
However, their difference in the tidal deformability is small and only exhibits in massive neutron stars, as shown in Fig.\ref{FIG2}(c).

\begin{figure*}[t]
\centering
\includegraphics[width=0.95\textwidth]{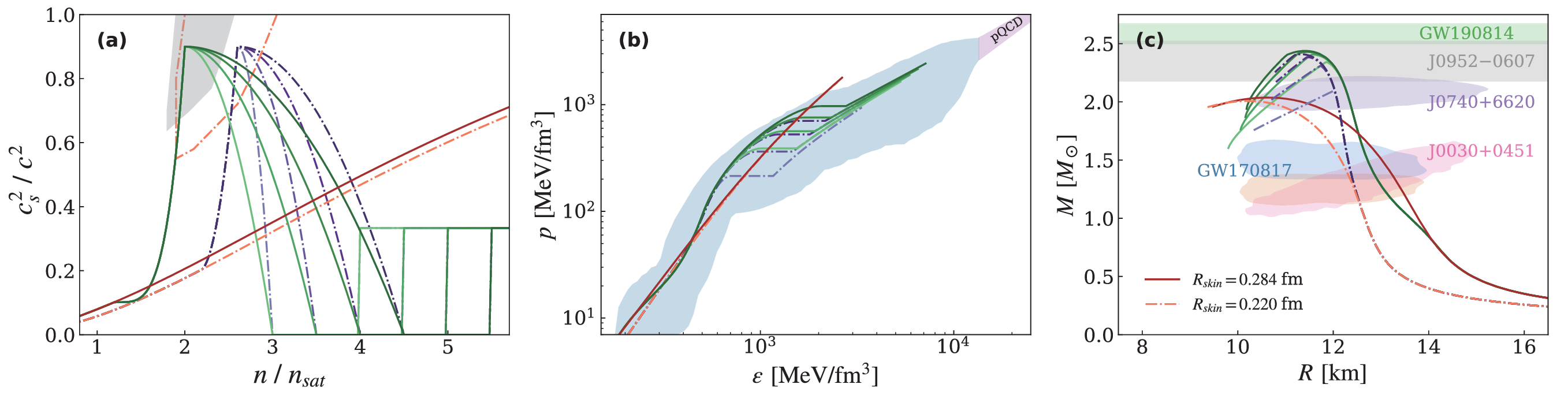}
\caption{
EoS and neutron star observations associated with different first-order phase transitions.
(a) the speed of sound with different first-order phase transitions, and the allowed peak regions associated with thick (shadow) and thin (orange dashed contour) neutron skins are displayed.
(b) the pressure corresponding to different sound speed in (a) and the allowed boundaries taken from Ref.~\cite{sound1};
(c) the mass-radius curves obtained with EoS in (b).
\label{FIG3}
}
\end{figure*}

The sharp first-order phase transition from nuclear matter to exotic quark matter is believed
to exist, e.g., as predicted to be around $\mu$$\thicksim$1200 MeV in the QCD phase diagram~\cite{LRP2015},
but it is a tremendous challenge to determine the actual transition point.
To explore the possible scenario of  phase transition, the consequences
of different transition points from 3.0 to 4.5$n_{sat}$ are compared in Fig.\ref{FIG3}.
The energy densities $\epsilon$ (or in chemical
potential $\mu$) at 3.0, 3.5, 4.0 and 4.5$n_{sat}$ are 920.2, 1215.9, 1547.2, 1913.2 MeV fm$^{-3}$, respectively.
In Fig.\ref{FIG3}, the starting density $n_1$ is fixed as 1.2$n_{sat}$, and
the peak location of $c_s^2$ is at 2.0$n_{sat}$.
For the thinner $R_{skin}^{208}$ of 0.22 fm, the allowed region of peak locations is much extended. 
The phase transitions with the thinner $R_{skin}^{208}$ are also shown for comparison, in which the peak is at 2.6$n_{sat}$.

The EoS associated with different speed of sound in Fig.\ref{FIG3}(a) and the resulting mass-radius curves are shown in Fig.\ref{FIG3}(b) and  Fig.\ref{FIG3}(c), respectively.
 For the thick neutron skin, the slight dip in  $c_s^2$ following 1.2$n_{sat}$
 causes a small softening deviation from the original EoS.
Such a small deviation of EoS is indispensable for describing neutron stars of 1.4$M_\odot$.
For the steep-rising or rapid-decreasing structures in the speed of sound,
the corresponding EoS changes rather moderately. 
Finally the EoS approaches the conformal limit of $c_s^2$=1/3. 
The allowed region of EoS from large-scale searches~\cite{sound1}  is also shown.
For the thick neutron skin, EoS related to the phase transition at densities $\geqslant4.5n_{sat}$ is reaching the allowed boundaries.
In addition, an early peak location with a delayed transition point is not consistent with a rapid drop of $c_s^2$ in a first-order phase transition~\cite{sound-3,fopt}. 
This indicates that 
an early first-order phase transition is more favorable with thick neutron skins. 
For the thin neutron skin, the allowed peak region is larger and extends to higher densities. 
Actually the combination of a delayed peak location and an early transition point
is problematic. For the transition point at 3.0$n_{sat}$ and the peak at 2.6$n_{sat}$, the maximum mass of 2.35$M_\odot$
can not be reached, as shown in Fig.\ref{FIG3}(c). 
This implies that a considerable coexisting region of hadrons and quarks before the first-order phase transition is needed to support 2.35$M_\odot$
for both thin and thick neutron skins. 
For the thin neutron skin, there is less preference in the transition point due to the extended peak area.  
The higher peak region associated with the thick neutron skin also means that the phase transition is stronger. 
Hence an early and strong first-order phase transition is comparatively more favorable for the thick neutron skin due to the stringent peak region. 
Note that an early first-order phase transition indicates the existence of large quark cores in neutron stars.

\emph{Discussions.}---
The present study is based on the astonishing thick neutron skin of $^{208}$Pb in experiments.
Note that there are large uncertainties in measurements~\cite{prex2}. Nevertheless
the lower limit of neutron skin thickness is still challenging for existing nuclear theories.
The first-principle calculations of nuclear matter above saturation density
are extremely difficult and uncertainties grow rapidly~\cite{sound-xft,CD}.
Ab initio nuclear calculations are actually based on renormalized two and three-body interactions in reduced configuration spaces~\cite{bshu,sound-xft,CD},
and omitted contributions of many-body short-range hard-core interactions could explosively increase at high densities.
In fact,  the average distance between nucleons is 1.46 fm at 2.0$n_{sat}$ in the cubic packing and nucleons are already strongly overlapped considering that the nucleon radius is $\thicksim$0.8 fm.
The threshold distance of 1.6 fm corresponds to 1.5$n_{sat}$.
This means current theoretical framework is questionable at 2.0$n_{sat}$ and the steep-rising structure in sound speed can not be excluded.
After the starting density $n_1$, EoS is required to  soften slightly and this could be due to the appearance of hyperons~\cite{hyperon}.
The softening EoS after the peak can be understood as the influences of deconfined quarks.
Very recently a possible light neutron star is observed with a mass of 0.77$M_\odot$ and a radius of 10.4 km~\cite{0.7m}.
If this object is a neutron star, then very early softening of speed of sound is required and  this
is not favorable for a thick neutron skin.
Therefore a more precise measurement of neutron skin is highly expected
because it has significant consequences.

\emph{Conclusion.}---
The surprising thick neutron skin
of $^{208}$Pb  is taken into account to constrain the speed of sound in neutron stars.
As a result, exotic structures in speed of sound are required, consisting of a small softening slope at the starting density $1.2\sim1.5n_{sat}$  followed by a steep rising peak around 2.0$n_{sat}$.
This exotic structures can successfully describe existing massive neutron stars and gravitational wave observations simultaneously.
Due to the stringent allowed region of peak locations in the case of thick neutron skins,
an early and strong first-order phase transition is more favorable. 
For the thin neutron skin, the allowed peak region is much extended and thus has less preference in the transition point.  
We reveal that the thick neutron skin has significant implications in 
 the speed of sound at intermediate densities, the appearance of hyperons and the first-order phase transitions, thus more precise
measurements of  $R_{skin}^{208}$ are highly expected.

\acknowledgments
We are grateful to discussions with L.W. Chen and F.R. Xu.
This work was supported by the National Natural Science Foundation of China under Grants
No.  11975032, 11835001, 12335007,  and 11961141003.
We also acknowledge the funding support from the State Key Laboratory of Nuclear Physics and Technology, Peking University (No. NPT2023ZX01).

\end{document}